\documentstyle[11pt,paspconf,epsf]{article}

\newcommand	{\B}[1]		{\mbox{\boldmath${#1}$}}
\renewcommand	{\bv} 		{\B{v}}
\newcommand	{\hbr}		{\hat{\B{r}}}

\newcommand	{\beq}		{\begin{equation}}
\newcommand	{\eeq}		{\end{equation}}
\newcommand	{\bef}		{\begin{figure}}
\newcommand	{\eef}		{\end{figure}}
\renewcommand	{\d}		{{\rm d}}
\def\lesssim{\mathrel{\hbox{\rlap{\hbox{\lower2pt\hbox{$\sim$}}}\raise2pt\hbox{$<$}}}}

\begin{document}

\title{The Distribution of Nearby Stars in Velocity Space}
\author{Walter Dehnen}
\affil{Theoretical Physics, Keble Road, Oxford OX1~3NP, U.K.}

\begin{abstract}
From the tangential velocities of stars all over the sky, one can, in a 
statistical way, infer their 3D velocity distribution. An application to 
Hipparcos data reveals rich structure in the planar stellar motions: there 
are several moving groups, increasing in number with stellar type but 
decreasing in importance. A distinct group of outward moving stars with 
low rotation velocities might be associated with the Galactic bar: growing 
a bar in a smooth model for the Galactic disk results in such a group, 
provided the Sun is outside the outer Lindblad resonance (OLR) and the 
orientation angle with the bar is in the first quadrant.

The vertical motions show less structure, but the mean $\bar{v}_z$
increases for large rotational velocities, which can be nicely explained
by a warp of the outer stellar disk.
\end{abstract}

\section{Inferring $f($\boldmath$v$\unboldmath$)$ from Hipparcos Data}
ESA's Hipparcos mission measured the positions, parallaxes, 
and proper motions of many nearby stars. The radial velocity (RV) of most of 
these stars is unknown (using only stars with known RVs gives
a sample strongly biased towards high-\bv\ stars [Binney et~al., 1997]),
and we cannot evaluate their space motions $\bv$, but only the tangential
velocities $\bv_t$. However, using stars from all over the sky, we can infer
their distribution $f(\bv)$ in a statistical way, for instance, by maximizing
the log-likelihood 
\beq \label{ll}
	{\cal L}(f)= N^{-1}\sum_i\ln\int\d v_r\,f(\bv=\bv_{t,i}+v_r \hbr_i)
\eeq
with $\hbr_i$ the unit vector in direction to star $i$. Since ${\cal L}(f)$ is 
unbounded, it must be regularized:\ ${\cal Q}_\alpha\equiv{\cal L}-\alpha{\cal
S}$ is maximized instead with ${\cal S}(f)$ a penalty functional measuring the 
roughness of $f(\bv)$ and $\alpha$ the smoothing parameter. The maximum of 
${\cal Q}_\alpha (f)$ subject to $f\,{>}\,0$ and $\int\!\d^3\!\bv f\,{=}1$ is 
the maximum penalized likelihood estimate (MPLE). The degree of smoothness, 
determined by $\alpha$, may be optimized by cross-validation using Monte-Carlo 
simulations (for more details and the numerics, see Dehnen, 1998, hereafter
D98).

This technique was applied to the kinematically unbiased sample of Hipparcos
stars defined by Dehnen \& Binney (1998, see also Binney, this volume). I 
analysed seperately four colour bins of main-sequence (MS) stars, the non-MS 
stars, and the full sample. For the latter, optimal smoothing was used, while
for the subsets, $f(\bv)$ was slightly undersmoothed to preserve features that
are likely to be real because they are present in more than one distinct subset.

Below I use a right-handed cartesian coordinate system with $\hat{e}_x,\,
\hat{e}_y,\,\hat{e}_z$ pointing towards $(\ell,b)=(0^\circ,0^\circ),\,
(90^\circ,0^\circ),$ and $b=90^\circ$.

\bef
\centerline{\epsfxsize65mm \epsfbox[42 315 582 716]{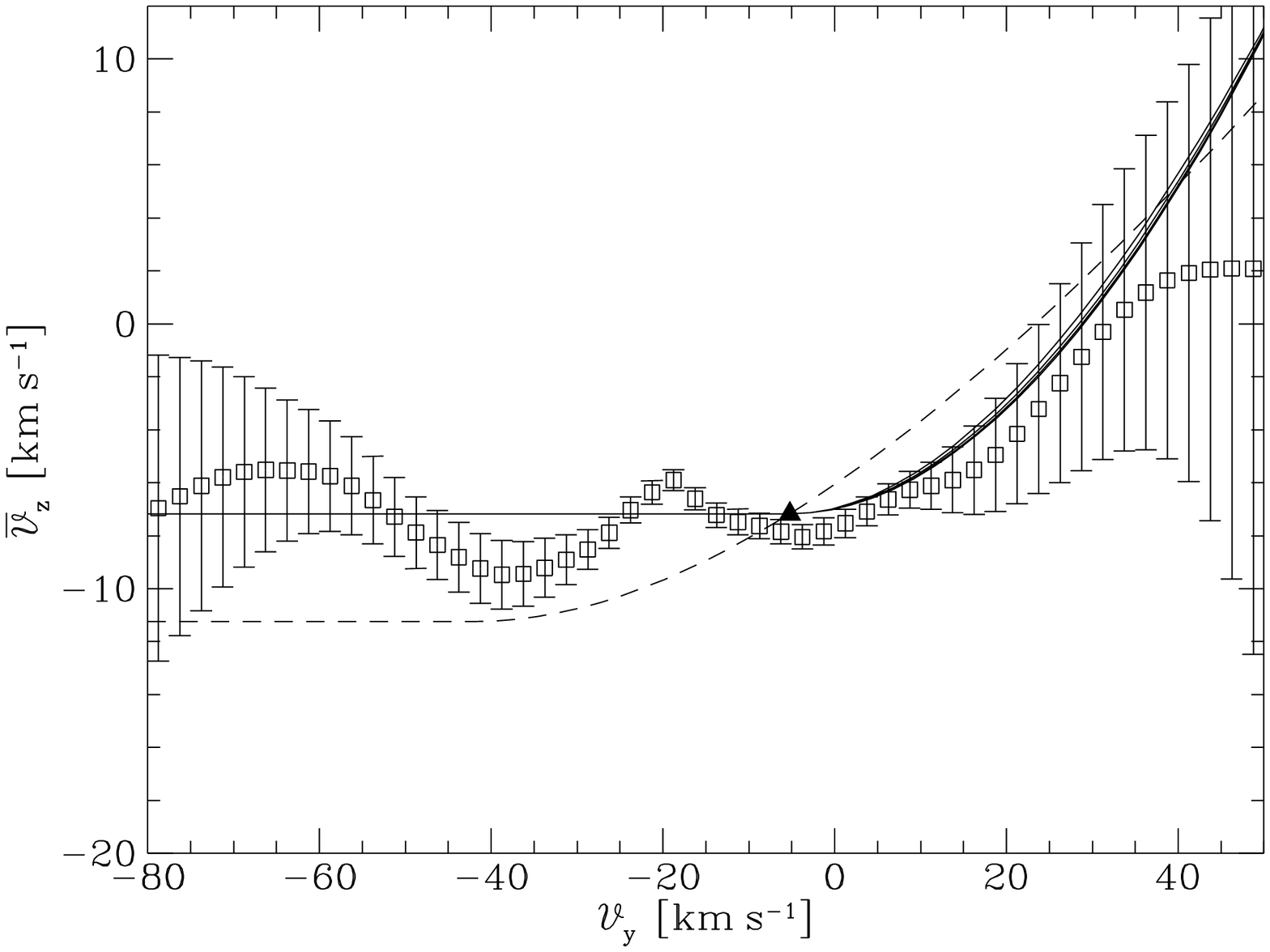}}
\caption{$\bar{v}_z$ measured for the full sample as function of $v_y$. The
$\triangle$ indicates the LSR. Lines represent models for the Galactic warp.}
\label{warp}
\eef

\section{Vertical Motions: Evidence for a Warp in the Stellar Disk}
The projections of $f(\bv)$ onto the $v_xv_z$ or $v_yv_z$ plane (Figs.~4\&5 in
D98) do not show any significant structure (see Binney, this volume, for 
reasons why). However, there is a skewness in the sense that $\bar{v}_z$ 
increases for large positive $v_y$, see Fig.~\ref{warp}. This effect results
naturally from a warp in the outer stellar disk:\ stars with large $v_y$ 
visit us from large $R$, where the warp causes them to move upwards.
The lines in Fig.~\ref{warp} are due to a simple quantitative model (see D98 
for details) for this effect. A warp starting within $R_0$ (dashed) cannot fit 
the data, while warps starting at $\sim R_0$ (solid) can. There is a degeneracy
between the warp's amplitude and pattern speed $\Omega_{\rm w}$ in the sense 
that 
\beq
	{\Omega_{\rm w}\over{\rm km\,s^{-1}\,kpc^{-1}}} \simeq
	{4-6(z_{10}/{\rm kpc})\over0.3(z_{10}/{\rm kpc})}
\eeq
where $z_{10}$ is the amplitude at $R\,{=}\,10\,$kpc. For H\,{\sc i}, $z_{10}
\simeq$0.3-0.4\,kpc, which would give $\Omega_{\rm w}\,{\simeq}\,$13 to 25\,$
\rm km\,s^{-1}\,kpc^{-1}$ (retrograde w.r.t.\ rotation).

\bef
\centerline{\epsfxsize134mm \epsfbox{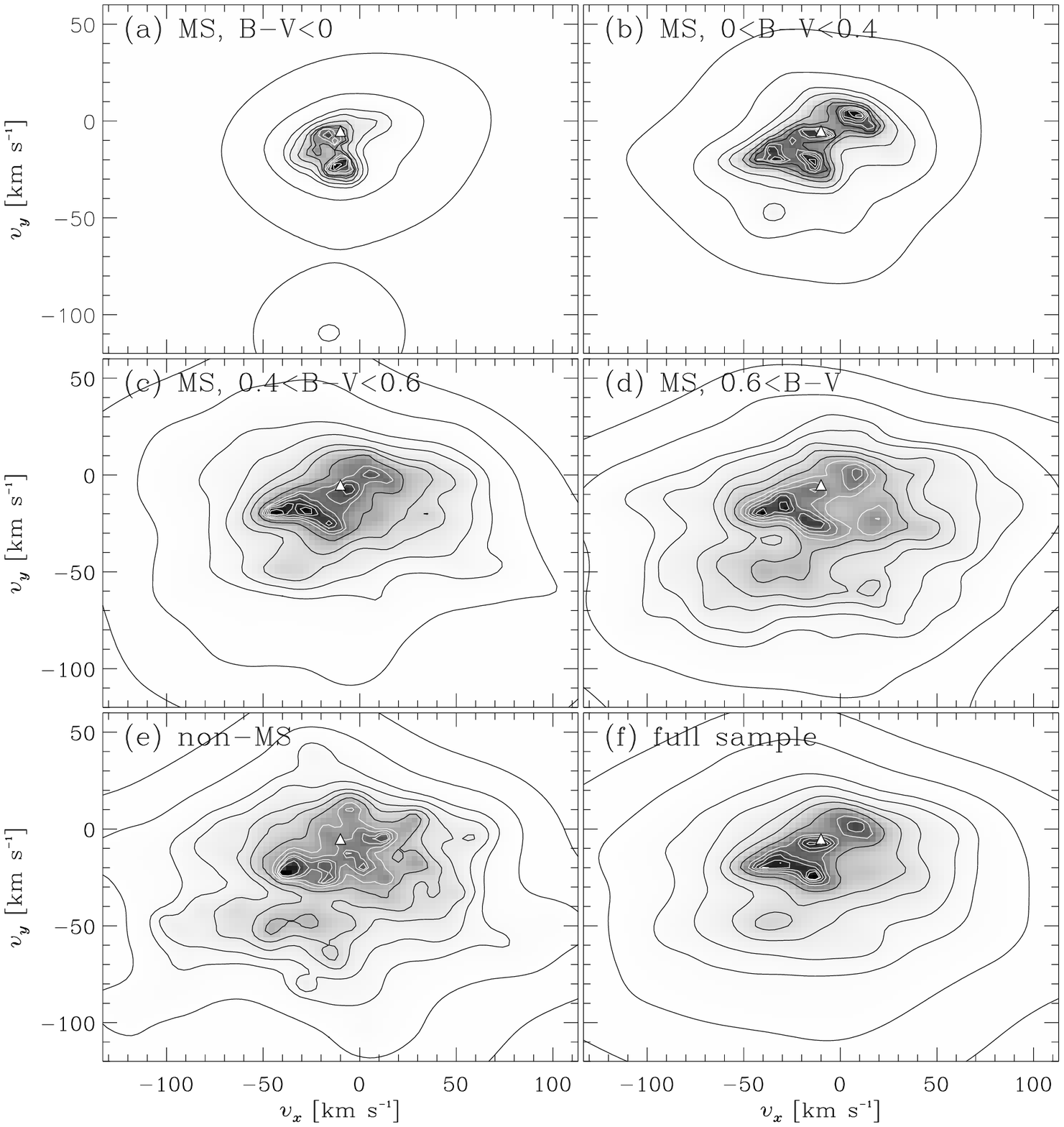}}
\caption[]{Distribution in planar velocities ($v_x$: towards $\ell=0^\circ$; 
	$v_y$: towards $\ell=90^\circ$) for the four colour bins of MS stars
	(a-d), the non-MS stars (e), and the full sample (f). Velocities are 
	relative to the Sun, and the $\triangle$ indicates the LSR. Contours 
	contain $2,\,6,\,12,\,21,\,33,\,50,\,68,\,80,\,90,\,95,\,99,\,$and 
	99.9\% of all stars. Only the full sample (f) is optimally smoothed, 
	the subsets are under-smoothed and contain artificial structures.
	However, features present in two or more subsets 
	are likely to be real, e.g.\ the double peak at $v_x{=}-25,-45\,\rm 
	km\,s^{-1}$ and $v_y{=}-50\,\rm km\,s^{-1}$ present in (d) and (e).\par
	Note the four major peaks, best visible in the upper right panel,
	which correspond to (from left to right) the Hyades, Pleiades, Coma,
	and Sirius moving groups. At later stellar types there are also many
	minor moving groups, predominantly at high velocities (w.r.t.\ the LSR)
	i.e.\ on eccentric orbits, but the total number of stars associated 
	with them diminishes with stellar type.
	}\label{fuv}
\eef

\section{Planar Motions: Moving Groups} 
The projection of $f$ onto the $v_xv_y$ plane shows significant structure, 
see Fig.~\ref{fuv}. Surprisingly, there are, apart from the well-known moving 
groups, also several minor features, mainly at high \bv, i.e.\ on eccentric 
orbits, and for late stellar types only, i.e.\ made of old stars. The standard 
picture of moving groups forming via dissolution and mixing of stellar clusters
cannot account for these features. A viable explanation, though, is a dynamical
formation of the moving groups, by means of resonant locking. If the Galactic 
potential changes slowly, the resonant islands slowly sweep through phase space.
Stars may then become locked to it and shifted to other phase-space regions
(cf.\ Sridhar \& Touma, 1996). In this picture, moving groups could be made of 
stars of different origin and age.

\bef[t]
\parbox[t]{67mm}{
        \epsfxsize=72mm \epsfbox[0 0 421 302]{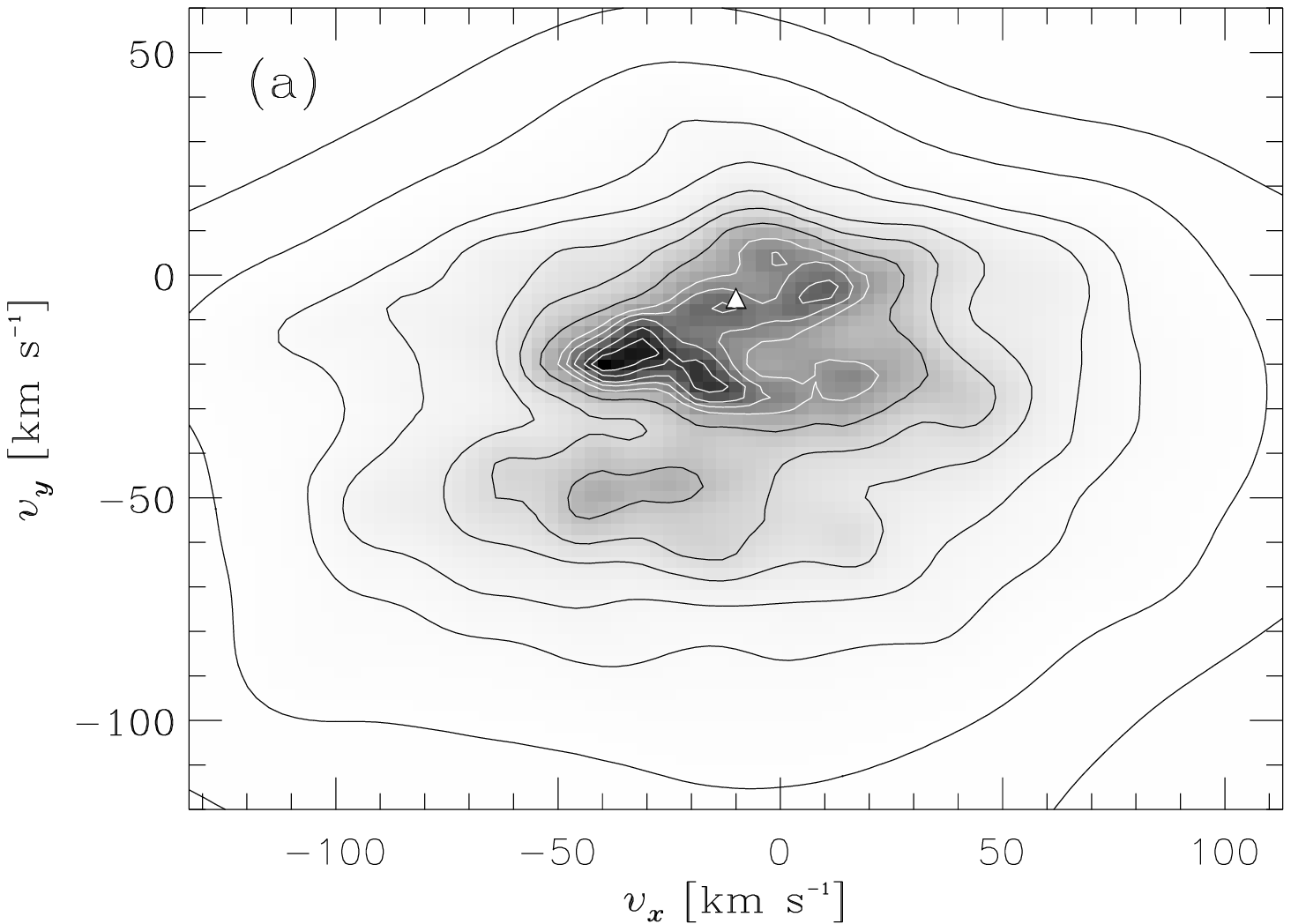}
} \hspace*{-12mm} \hfill
\parbox[b]{72mm}{
\caption[]{\sloppy Top: as Fig.~\ref{fuv} but (d) and (e) analyzed together. 
Bottom: $f(\bv)$ obtained from modeling: a bar has been grown in an initially 
smooth, axisymmetric, and exponential disk. The co-rotation radius is 4kpc, 
while solar radius and bar position angle are
$(R_0,\phi)\,{=}\,(7.5\,{\rm kpc}, 10^\circ)$ for (b) and
$(8.0\,{\rm kpc}, 40^\circ)$ for (c). } 
\vspace*{8mm} \label{bar} } 
\par
\centerline{ \epsfxsize=72.08mm \epsfbox[3  0 421 320]{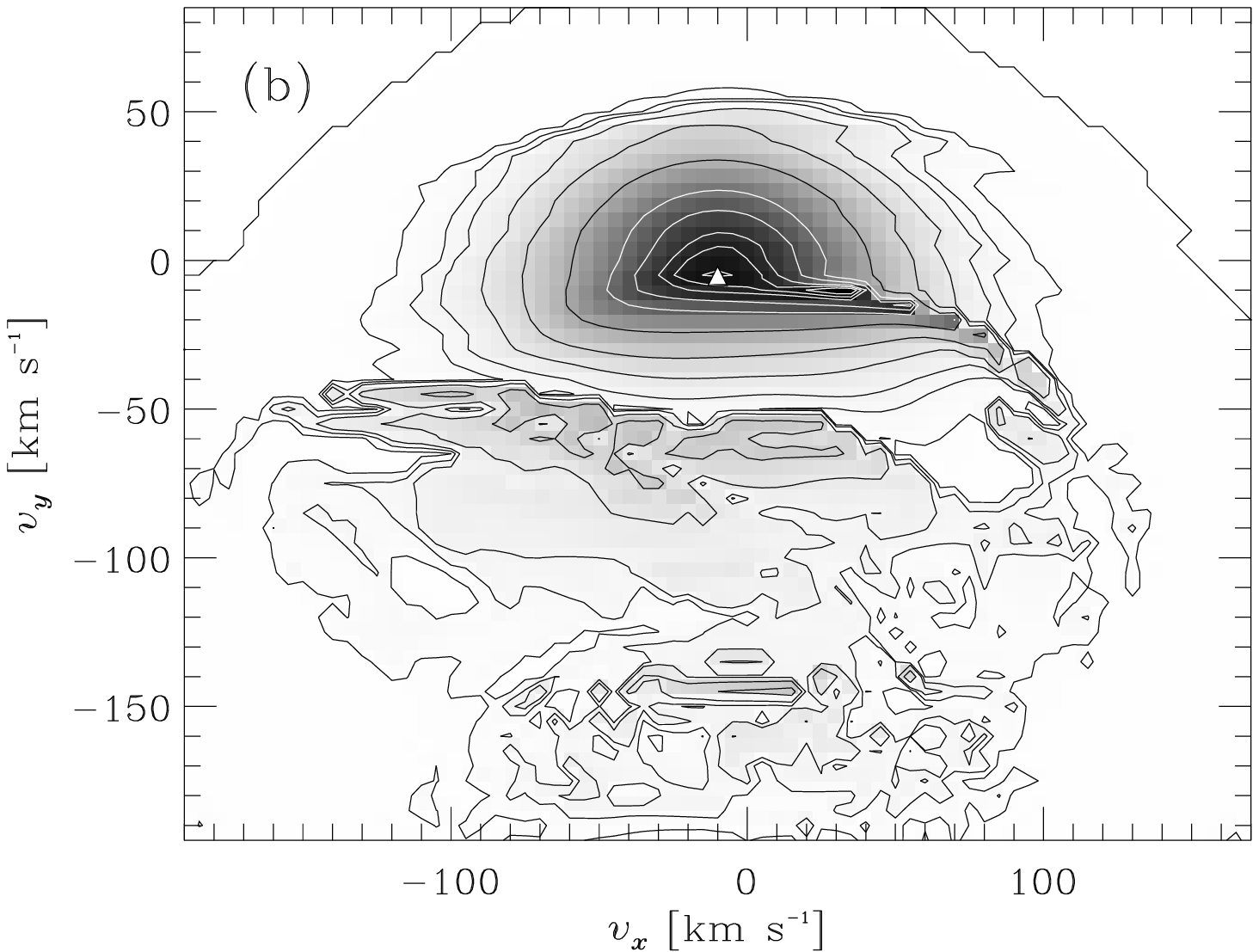} \hfill
             \epsfxsize=61.92mm \epsfbox[62 0 421 320]{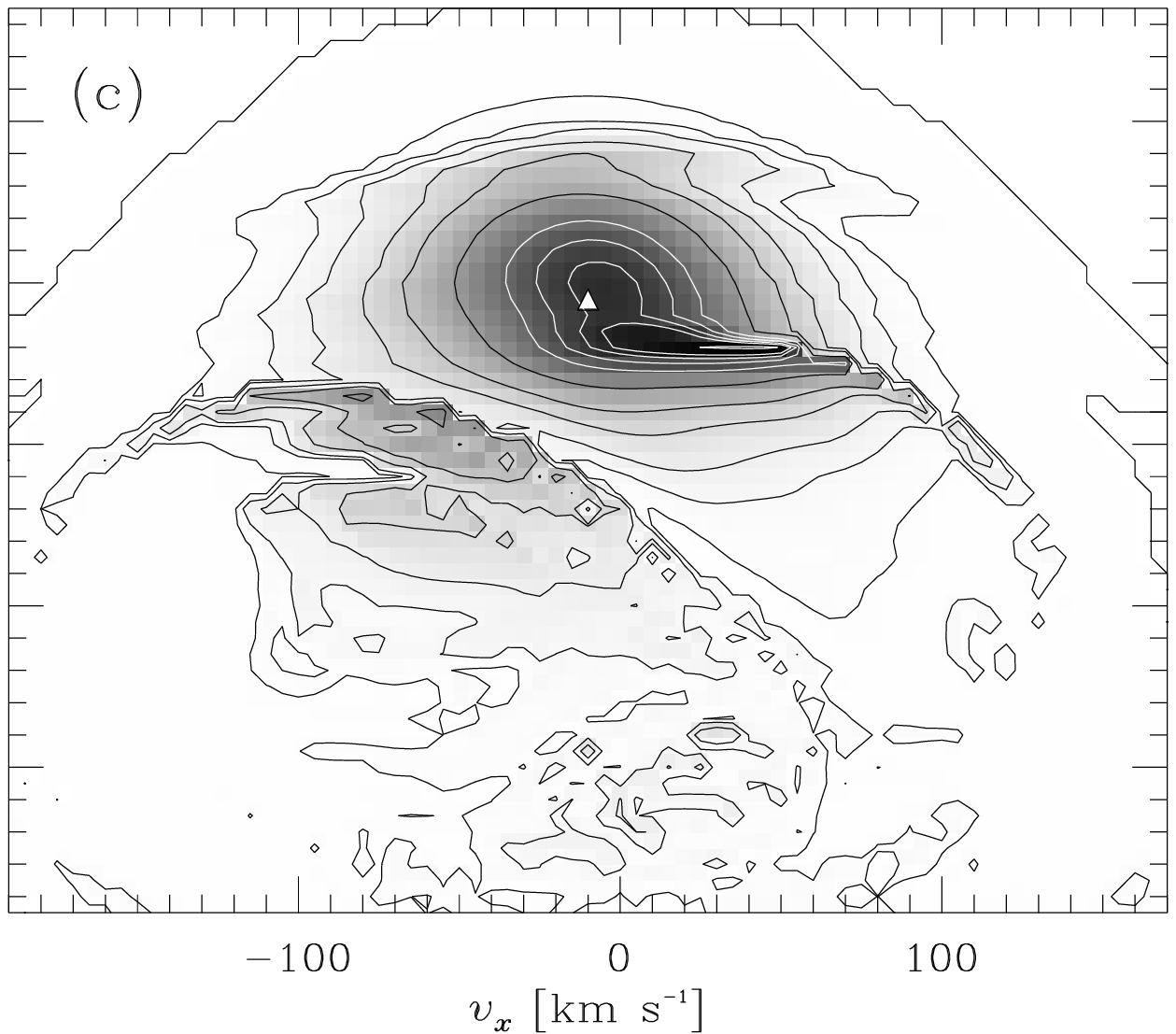} }
\eef

\section{Planar Motions: Influence from the Galactic Bar?} 
There is a distinct group of stars at $v_y{\simeq}-50$ and $-100\lesssim v_x
\lesssim20$ (in km\,s$^{-1}$), see Fig.~\ref{bar}a, yielding $\bar{v}_x{<}0$
at $v_y{<}{-}30\,\rm km\,s^{-1}$, also found by Raboud et~al.\ (1998) from a 
sample of Hipparcos stars with known RVs. These authors report high $Z$ for
these stars and argue they are a signature of the Galactic bar, as
in Fux's (1997) models $\bar{v}_x{<}\,0$ at bar position angles $\phi\,{\in}\,
(0^\circ,90^\circ)$. 
Such a feature is created naturally by a bar grown in an initially axisymmetric
disk, Fig.~\ref{bar}b\&c, if we are outside the OLR. In this scenario, the 
outward moving stars visit us from inside the OLR, where the orbits move 
outwards for $\phi\,{\in}\,(0^\circ, 90^\circ)$. The position of this group
in $\bv$-space depends sensitively on $\phi$ and our distance to the OLR.

\end{document}